**From Architectural Sketch to Conceptual Representation**

*Using structure-aware diffusion model to generate renderings of school buildings*

Zhengyang Wang[1], Hao Jin[1], Xusheng Du[1], Yuxiao Ren[2], Ye Zhang[2] and Haoran Xie[1]
[1]*Japan Advanced Institute of Science and Technology*
[2]*Tianjin University*
[1]*s2310021@jaist.ac.jp, ORCID: 0009-0008-6600-6911*

**Abstract.** Generative Artificial Intelligence (AI) has advanced rapidly, enabling the generation of renderings from architectural sketches. This progress has significantly improved the efficiency of communication and conceptual expression during the early stage of architectural design. However, generated images often lack the structural details from architects' sketches. While sketches typically emphasize the overall structure, crucial components such as windows and doors are often represented by simple lines or omitted entirely. For school buildings, it is essential to control architectural components, such as the shape and proportion of windows, as these factors directly influence the accuracy of the generated images in reflecting the architect's design intentions. To address this issue, we propose a structure-aware diffusion model for architectural image generation to refine expressing design intentions through retrieval augmentation. Our framework utilizes architectural components to enhance the generation process, addressing the details that may be lacking in the sketches. These components provide clear spatial and structural details, improving the model's ability to interpret and generate architectural details. The refined sketches, combined with text prompts, are fed into the proposed structure-aware diffusion model to generate detailed and realistic school building images. The experiment results demonstrate the effectiveness of our framework in generating architectural designs.

## 1. Introduction

During the early stages of architectural design, designers frequently engage in the iterative process of creating renderings to communicate design concepts and architectural intentions to stakeholders. Compared to the traditional tools such as Sketchup, Rhino and ArcGIS, image generation models like Midjourney can significantly reduce the time and effort required for image production, making the process more efficient (Li et al., 2024). However, the current text-to-image generation models face challenges in controlling the structure of buildings in image generation. Text prompts alone often fail to convey specific details such as window placements, floor counts, or proportional relationships between structural architectural components.

In this work, we especially focus on school buildings, which are architectural structures with typically featured designs and detailed architectural characteristics. Generating images of school buildings through text-to-image models remains a challenge (Paananen et al., 2024), as the text input may have difficulty controlling the details of the building. This limitation impedes architects in effectively communicating their design intentions, reducing the clarity and precision of the architectural concept presented to clients and stakeholders. Meanwhile, for school buildings, the window-to-wall ratio can significantly influence factors such as natural lighting, energy efficiency, and the attention levels of students, making precise architectural detailing even more critical (Heschong et al., 2002). To address this issue, we adopt sketches as an additional condition, providing structural foundation information such as the overall structure and proportions. Therefore, we choose to generate architectural images of school buildings from user sketches combined with text prompts and focus on depicting architectural components like windows and doors. When user-provided sketches show the overall structure and layout but are often simplified and lack details like window placement, it becomes difficult for generative models to produce accurate and realistic architectural images. Detailed sketches with structural details provide clearer and more precise guidance to facilitate the generative process. To achieve this, architectural component regions are extracted and then matched with architectural components from an architectural component dataset using a sketch retrieval approach. The retrieved components are then integrated into the sketches to add details like windows and doors. As a result, the generated architectural images that are both detailed and aligned with traditional architectural designs generative approaches.

The main contributions of this work are listed as follows:

- We propose a novel approach that leverages user sketches as input, enabling precise generation of architectural designs with structural details.

- We establish two datasets: one for retrieving augmented components and the other for pairing sketches with their corresponding high-quality renderings, providing robust support for precise and realistic generation.

- We demonstrate that incorporating retrieval-augmented generation with an architectural component dataset significantly improves the accuracy and realism of generated architectural designs, ensuring alignment with design intentions.

## 2. Related Works

### 2.1. DIFFUSION MODEL

Diffusion models have been widely adopted for image generation tasks, demonstrating their capability to generate high-quality and diverse images. Latent Diffusion Model (LDM) (Rombach et al., 2022) is a variant of the diffusion model. It consists of an encoder, a decoder, and a denoising U-Net network. Different from Denoising

Diffusion Probabilistic Models (DDPM) (Ho et al., 2020) that operates directly in the pixel space, the LDM's encoder compresses images into latent space, while the decoder reconstructs the images from it. Within the latent space, the forward diffusion process gradually adds noise into latent representations at each time step. The reverse diffusion process is the inference process of denoising. By compressing images from pixel space to latent space, LDM significantly decreases computational complexity, enabling faster sampling and more efficient training. Conditional diffusion models, like ControlNet (Zhang et al., 2023), enhance control over generated content by introducing conditions. These models have been widely explored in architectural design for residential floor plan generation (Shabani et al., 2023), and shear wall structure generation (He et al., 2023). These models can employ various input conditions (e.g., depth maps and segmentation maps) to generate images, aligning the generated image more closely with user requirements. However, these models still face significant challenges when generating architectural designs. Specifically, during the early stages of architectural design, obtaining inputs like depth maps and segmentation maps is difficult, which limits the applicability of these models for initial concept design. To address this limitation, our work focuses on generating conceptual representations of school buildings directly from architectural sketches.

## 2.2. RETRIEVAL-AUGMENTED GENERATION

Retrieval-Augmented Generation (RAG) (Lewis et al., 2020) provides an efficient approach to guide and improve the generation process using relevant information from an external database without training network parameters. In image generation tasks, RAG enhances the performance of generative models by improving the quality and accuracy of image synthesis (Sheynin et al., 2022; Golatkar et al., 2024). The core of RAG is its ability to extract relevant information from a large external database, guiding the generation process and greatly enhancing both the quality and accuracy of image generation. Our work employs RAG to convert a rough input sketch into a detailed sketch. The RAG process enhances rough input sketches by retrieving and integrating architectural components, such as windows and doors, ensuring realistic representations during the conceptual design stage.

## 3. Method

In this paper, we propose a sketch-guided framework of structure-aware diffusion model with retrieval augmentation for generating high-quality architectural design of school buildings directly from conceptual sketches during the early design stage. Figure 1 illustrates the workflow of our framework. Initially, the rough input sketch undergoes segmentation to extract architectural components (e.g., windows and doors). Refined architectural components are then retrieved from the architectural component dataset, used to generate a detailed sketch. Finally, the text prompts and the detailed sketch serve as conditioning inputs for the latent diffusion model to generate a high-quality school building image.

## 3.1. RETRIEVAL AUGMENTATION

### 3.1.1. Architectural Component Extraction

Segment Anything Model (SAM) (Kirillov et al., 2023) extracts precise segmentation masks guided by various inputs, such as text prompts, bounding boxes, and points. In our framework, SAM is utilized to segment architectural components such as windows and doors from rough input sketches. When the sketch contains ambiguous or incomplete regions, we adopt a semi-automatic approach, leveraging manual input (e.g., adding points or bounding boxes) to guide the segmentation process. SAM consists of three components: the image encoder, the prompt encoder, and the mask decoder. The image encoder and the prompt encoder map input images and text into corresponding feature space representations. Subsequently, the mask decoder combines the embeddings from the image encoder and the prompt encoder, decoding the feature map to produce segmentation masks.

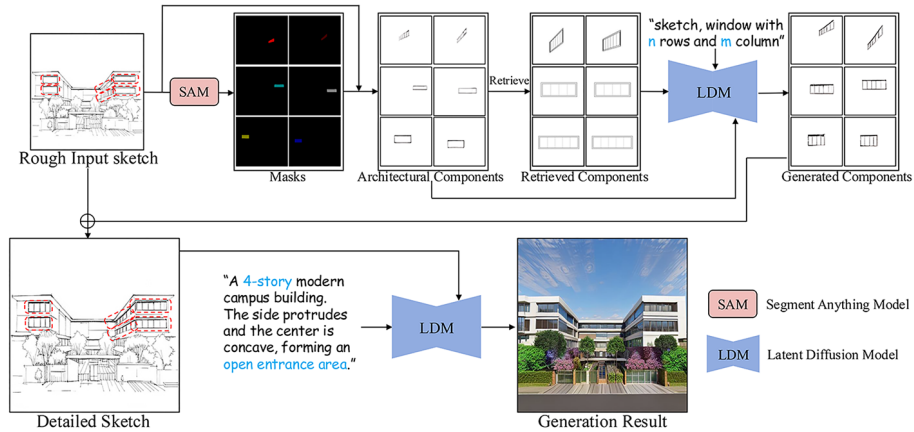

Figure 1. An overview of our proposed framework. Given a rough input sketch, users retrieve desired components (e.g., "a window with 1 row and 4 columns") through SAM approach and RAG approach, then fed to LDM for producing a detailed sketch. The detailed sketch and text prompt are then used in the generation LDM to generate the final architectural image.

### 3.1.2. Structural Retrieval and Refinement

Sketch-based image retrieval (Eitz et al., 2012) aims to calculate the similarity between an input image and images in an external dataset. The retrieval method normally uses the Bag of Features to encode line segments, extracts feature histograms and constructs a visual vocabulary. By extracting features from the input image, sketch-based image retrieval can efficiently retrieve similar images from the dataset based on these features. This method was employed to retrieve suitable components that best match the sketch from the architectural component dataset, based on the components segmented from the rough sketch input.

In addition, we adopt a LDM to refine the rough input sketch. A mask is applied to cover the rough input sketch to target the regions that need refinement. A cross-attention layer for sketch features then employed to effectively combine the masked rough input sketch and the text prompt. Furthermore, ControlNet is used as a component encoder to extract the features from the retrieved architectural components, which are then injected into the LDM. Finally, the generated architectural components that align with the rough input sketch are mapped back onto the unmasked regions of the rough input sketch, resulting in a detailed sketch. In this way, our framework can transform a rough input sketch into a detailed sketch by incorporating structural information from the retrieved components.

## 3.2. STRUCTURE-AWARE GENERATION

In this work, we propose a structure control mechanism aimed at ensuring the structural accuracy and architectural consistency of generated images, with the goal of achieving realistic building image generation. We train a diffusion-based model integrated with a building encoder, which generates school building images using detailed sketches and text prompts as input conditions. The building encoder leverages the detailed sketch to capture architectural features such as window locations, the number of floors, elevated structure, and suspended structure. Additionally, the text prompts not only describe the building's structure but also provide supplementary details like facade materials and architectural styles, enhancing the design and ensuring a more comprehensive architectural design. Figure 2 showcases some of architectural designs generated using the proposed framework.

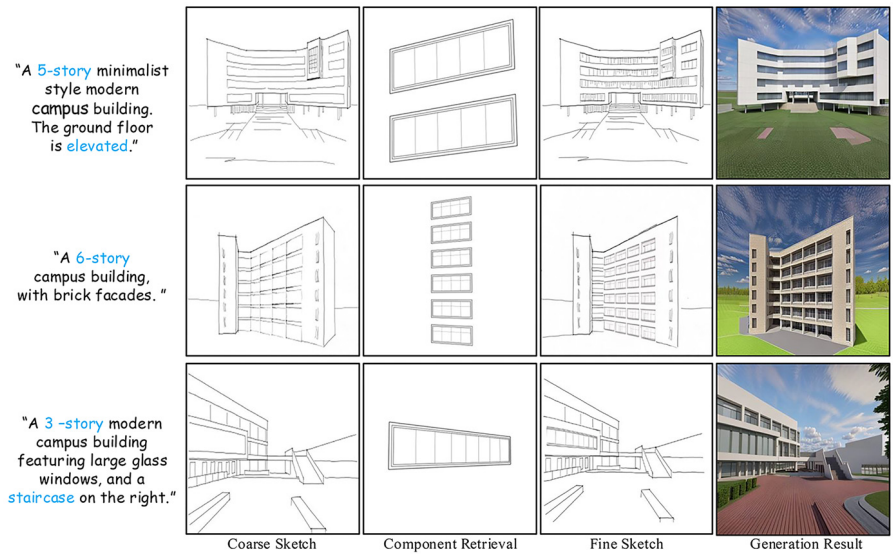

Figure 2. The generation process of architectural designs.

The training process begins with an input latent representation image $x_0$. At time step t, noise is incrementally adding noise $\epsilon$ to the image $x_0$, producing a noisy image $x_t$. To train the model to restore the noisy image $x_t$, the model learns to estimate the added noise. By learning to predict the noise added at each time step t, the model can reverse the process during generation to restore a noisy image $x_t$ to the image $x_0$. The loss function is given as follows:

$$\mathcal{L} = \mathbb{E}_{x_0,t,c_t,c_s,\epsilon \sim \mathcal{N}(0,1)}[\|\epsilon - \epsilon_\theta(x_t, t, c_t, c_s)\|_2^2],$$

where $C_t$, $C_s$ are text prompts and the detailed sketch. $\epsilon_\theta(x_t, t, c_t, c_s)$ is the noise predicted by the model given the noisy image $x_t$, time step $t$, text prompts $c_t$, and the detailed sketch $c_s$.

## 4. Implementation Details

### 4.1. ARCHITECTURAL COMPONENT DATASET

To address the issue of simplified, incomplete, or ambiguous regions in sketches during the early stages of architectural design, we established an architectural component dataset comprising 300 images of architectural components. The dataset includes various arrangements of architectural components, such as windows and doors. Figure 3 illustrates different arrangements of architectural component, including variations in scale, perspective, and orientation. These architectural components are carefully categorized by type. By incorporating flexible architectural component arrangements, the approach ensures that the diversity in the number and arrangement of architectural components minimizes errors in generating building structures due to mismatched component counts.

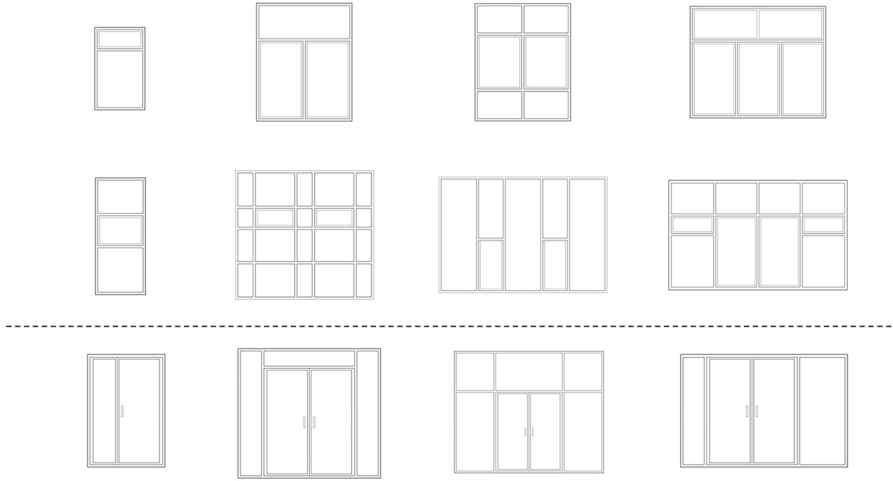

Figure 3. Example images from the architectural component dataset. These illustrate architectural components representing windows and doors which are utilized in the retrieval process to refine rough input sketches.

### 4.2. TRAINING DATASET

To generate final renderings from detailed sketches, we established a specialized dataset for training. As shown in Figure 4, this dataset consists of 1,600 pairs of detailed sketches, text prompts, and final renderings of school buildings. In the dataset preprocessing phase, we employed multiple approaches to improve the quality and diversity of the dataset. Initially, we employed ChatGPT (GPT-4) to automatically generate text prompts. These text prompts are then manually corrected by experts to ensure accuracy. Subsequently, we utilized the sketch style transfer method (Chan et al., 2022) to transform sketches into diverse styles, enhancing the model's adaptability. We use this method to diversify the data with different representation styles. Furthermore, we integrated data augmentation techniques such as horizontal flipping and random cropping to diversify the dataset further and enhance the model's robustness and adaptability to different input school buildings.

## 4.3. TRAINING DETAILS

In our training stage, our experiments were conducted using a high-performance single NVIDIA GeForce RTX 4090 GPU with 24 GB of memory. The training model used PyTorch version 2.5.1. The training parameters involved a total of 400,000 training steps with a batch size of 4, and a learning rate of 1e-5. These pairs of sketches and their corresponding rendered images were resized to a resolution of 512x512 to ensure uniformity across the dataset and maintain compatibility with the pretrained model. To reduce the number of training iterations and better adapt to the scale of our dataset, the pretrained weight of stable-diffusion-v1-5 model was used as the base.

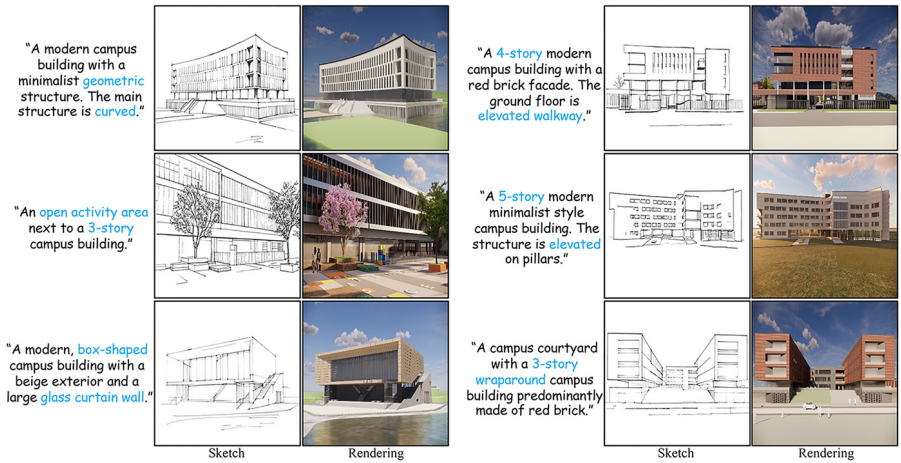

Figure 4. Example images from the building dataset. Our dataset contains text prompts, sketches, and the corresponding renderings.

## 5. Evaluation

### 5.1. QUANTITATIVE STUDY

To evaluate the performance of our proposed framework, we compared it with the state-of-the-art diffusion models, including Stable Diffusion (SD), Stable Diffusion XL (SDXL) (Podell et al., 2023), and ControlNet (Zhang et al., 2023), using their official pre-trained weights without training. Figure 5 showcases the comparison of the generated architectural design results. It is important to note that during the early stages of architectural design, there is often no definitive "true" design solution, as this phase is characterized by the exploration of multiple possibilities. Our method does not aim to generate a single "correct" output but rather focuses on producing renderings that align with professional design practices and architectural principles.

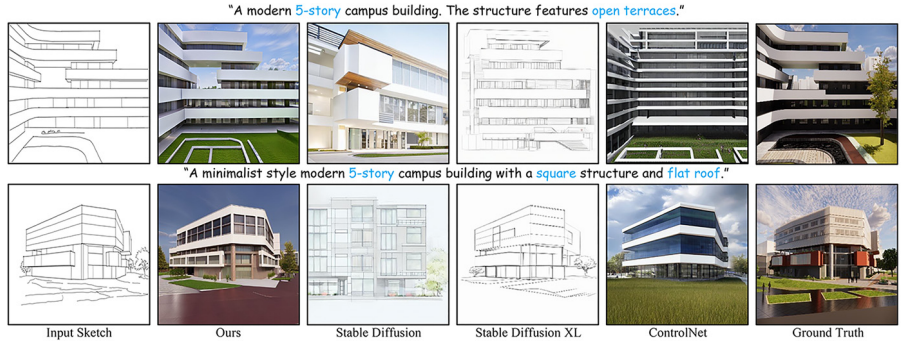

Figure 5. Visual comparison of building images generated by Stable Diffusion, Stable Diffusion XL, ControlNet, and our proposed framework, using the same text prompts and sketches as input.

We utilized evaluation metrics such as Peak Signal-to-Noise Ratio (PSNR) (Huynh-Thu and Ghanbari, 2008), Structural Similarity Index (SSIM) (Wang et al., 2004), and Learned Perceptual Image Patch Similarity (LPIPS) (Zhang et al., 2018) to quantitatively evaluate the quality of the generated images. As shown in Table 1, the PSNR values demonstrate that the differences between our model's generated images and the ground truth images in terms of RGB pixels are smaller than those of the other models. And the SSIM values, which are higher than those of other models confirm that our model generates architectural designs that come closer to real images in terms of luminance, contrast, and structural information. The PSNR and SSIM values corroborate the superior quality of our generated images in terms of detail preservation and overall clarity. Furthermore, the LPIPS metric, which captures high-dimensional image features and perceptual differences, is particularly effective for evaluating alignment with real-world scenarios. The lowest LPIPS values indicate that our model's generated images closely approximate the ground truth images in visual characteristics and aesthetic appearance.

Table 1. Quantitative comparison of image generation performance across different models, including SD-1.5, SDXL-1.0, ControlNet, and our proposed framework.

| Model | PSNR↑ | SSIM↑ | LPIPS↓ |
| --- | --- | --- | --- |
| SD-1.5 | 7.033 | 0.22 | 0.70 |
| SDXL-1.0 | 5.073 | 0.29 | 0.68 |
| ControlNet | 9.132 | 0.18 | 0.60 |
| Ours | **16.19** | **0.59** | **0.24** |

5.2. USER STUDY

To further evaluate the generation capabilities of our proposed framework, we conducted a user study involving a total of 70 participants, graduate students specializing in architecture and design. These participants provided feedback and assessments on the quality of the images generated by the framework. Participants were asked to evaluate the generated images based on three questions: the quality of the generated images, their alignment with the input sketches, and the accuracy of architectural details in the generated images. We used a 5-point Likert scale to evaluate the results. In terms of the quality of the generated images, the mean value (MEAN) and the standard deviation (SD) are 3.93 and 0.98 respectively, which indicates that the generated images were well-received. The sketch alignment received a high rating with a mean value of 4.03 (SD = 0.47), which suggests a high level of participant satisfaction with the consistency be the generated images and the input sketches. The detailed accuracy also garnered substantial favourable reviews, achieving an average rating of 3.97 (SD = 0.48), indicating the generated details were generally considered accurate.

Through interviews with participants, we gained a deeper understanding of their evaluation. Some participants noted that the model effectively generated an accurate number of floors based on input sketches. Others suggested potential improvements, such as extending the framework to support more complex buildings or even entire building complexes, to broaden its applicability.

**6. Conclusion**

In this paper, we proposed a sketch-guided framework with retrieval augmentation to generate architectural designs with specified components for school buildings. By leveraging diffusion models, our framework effectively integrated rough input sketches and retrieved architectural components to produce architectural designs. Our approach addressed the challenges faced by existing diffusion models in terms of controlling architectural components in the generated architectural designs, while also enabling faster communication of design concepts and architectural intentions to all stakeholders during the early stages of the design process.

This work also has several limitations. Firstly, while our proposed framework enables the replacement of localized areas in the sketch, the regenerated areas may not be well integrated with the overall design, which may affect the coherence of the generated results. For example, when there are trees, pillars, and other obstructions around the area to be modified in the sketch, the generated result may not match the shape of these obstructions. This misalignment can make the architectural design appear visually unnatural or inconsistent. Additionally, Our framework lacks adaptability to extreme scenarios, such as very small spaces or complex terrain architectures. It is difficult to generate high-quality results under these conditions. Furthermore, this study does not address architectural components beyond doors and windows. We would like to enhance the diversity of the database with various building types and consider improving local inpainting techniques to ensure better integration of regenerated areas with the overall design as the future work.